\documentclass{revtex4}
\usepackage{graphicx}
\usepackage{fancyhdr}
\usepackage{amsmath}
\usepackage{color}
\usepackage[T1]{fontenc}
\newcommand{\ee}{\end{equation}}
\newcommand{\be}{\begin{equation}}
\newcommand{\ba}{\begin{array}}
\newcommand{\ea}{\end{array}}
\newcommand{\m}{M_{H^{\pm}}}
\newcommand{\g}{\,\mbox{GeV}}
\newcommand{\la}{\lambda_1}
\newcommand{\lb}{\lambda_2}
\newcommand{\lc}{\lambda_3}

\newcommand{\lczp}{\lambda_{345}}
\newcommand{\rg}{R_{\gamma\gamma}}

\newcommand{\fr}{\frac}
\usepackage{amssymb}
\usepackage{amsthm}
\usepackage{url}
\usepackage{verbatim}
\usepackage[lofdepth,lotdepth]{subfig}
\begin{document}

\title{Inert Doublet Model with a 125 GeV Higgs}

%

\author{M. Krawczyk, D. Soko\l owska, B. \'Swie\.zewska}
\affiliation{University of Warsaw, Faculty of Physics}

\begin{abstract}
 A 125 GeV Higgs-like particle discovered at the LHC in 2012 has properties expected for it in the Standard Model (SM), with a possible enhancement in the two-photon channel. Such SM-like Higgs scenario can be realized within the  Inert Doublet Model  (IDM) - a version of the Two Higgs Doublet Model with an exact discrete $D$ ($Z_2$-type)  symmetry.  In this model one SU(2) doublet plays the role of the SM Higgs doublet with one SM-like Higgs boson.  The second doublet has no vacuum expectation value and does not interact with fermions. Among four scalars constituting this $D$-odd doublet the lightest one is stable,  being if neutral a  good DM candidate  with the right  relic density. In this paper an analysis of the two-photon Higgs decay rate in IDM,  respecting  theoretical and other experimental  constraints, is presented. The enhancement in $h\to \gamma \gamma$ is possible only if invisible channels are closed,  with the enhancement  $R_{\gamma \gamma}>$1.2 for  masses of DM and charged scalars  below 154 GeV.   The temperature  evolution of the Universe to the Inert phase (described by the IDM)  in  $T^2$ approximation is presented and all possible sequences of phase transitions 
in one, two and three steps are analyzed. Going beyond this approximation by using an effective potential approach with one-loop $T=0$  Weinberg-Coleman term and temperature dependent effective potential at $T\not =0$ we address the question, whether the strong first-order phase transition needed for baryogenesis can be realized in IDM. A region with a good DM relic density and a strong first-order phase transition  is established, and discussed in the light of the XENON-100 data. 

\end{abstract}

\maketitle

\thispagestyle{fancy}


\section{Introduction \label{intro}}
The Theory of Matter refers to the basic concepts like the quantum field theory framework, gauge symmetry and the assignments of the lightest fermions. The idea of Spontaneous Symmetry Breaking (SSB) also  has such status, although 
a particular realization of the SSB  is not known and various models are considered.  
Standard Model (SM) is based on  one SU(2) doublet of spin-zero fields. Standard extensions of the SM are models with two doublets (2HDM), which can be realized in MSSM.  Similarly, models with more doublets, as well as with singlets, can be considered, all leading to  $\rho=1$ at the tree-level. In 2HDM various versions of the Yukawa interactions are possible.  

There are five scalars in 2HDM, among them neutral SM-like Higgs particles. The new particle, discovered in 2012 at the LHC,  strongly resembles the Higgs boson from the SM.  Here we consider a particular version of 2HDM - the Inert Doublet Model (IDM), which is very similar to the SM, in the sense that only  one SU(2) doublet
is involved in the SSB and there exists only one Higgs particle with the tree-level properties  like the Higgs boson
in the SM, $H_{SM}$.  The properties of the second doublet are quite different: it is not involved in the SSB and does not interact with fermions. It  contains four dark scalars, which have limited interactions with the SM particles and the lightest of them is stable, thus, if neutral, being a good candidate for the Dark Matter (DM) particle. 
\section{$D$-symmetric 2HDM\label{model}}
The real content of the theory is determined by  the symmetry properties of the Lagrangian as well as of the vacuum state. 
Here we assume that the potential is $Z_2$-symmetric with respect to the transformation $\phi_S\to \phi_S,\,\,\, \phi_D\to -\phi_D$, which we call  below the $D$ symmetry.  This symmetry leads to the CP conservation in the model.

The $D$-symmetric  potential has the following form:
\begin{displaymath}\begin{array}{c}
V=-\fr{1}{2}\left[m_{11}^2(\phi_S^\dagger\phi_S)\!+\! m_{22}^2(\phi_D^\dagger\phi_D)\right]+
\fr{\lambda_1}{2}(\phi_S^\dagger\phi_S)^2\! 
+\!\fr{\lambda_2}{2}(\phi_D^\dagger\phi_D)^2\\[2mm]+\!\lambda_3(\phi_S^\dagger\phi_S)(\phi_D^\dagger\phi_D)\!
\!+\!\lambda_4(\phi_S^\dagger\phi_D)(\phi_D^\dagger\phi_S) +\fr{\lambda_5}{2}\left[(\phi_S^\dagger\phi_D)^2\!
+\!(\phi_D^\dagger\phi_S)^2\right],
\end{array}\label{pot}\end{displaymath}
with all  parameters real. We take $\lambda_5<0$ without loss of generality \cite{Krawczyk:2010}. 

Various extrema (which can be local or global minima or the saddle points) can be realized in this potential. The vacuum expectation values (vevs) are as follows:
\begin{equation}
\langle\phi_S\rangle =\frac{1}{\sqrt{2}} \begin{pmatrix}0\\ v_S\end{pmatrix}\,,\qquad \langle\phi_D\rangle = \frac{1}{\sqrt{2}}
\begin{pmatrix} u \\ v_D  \end{pmatrix}, \quad v_S, v_D, u \in \mathbb{R}. \label{dekomp_pol}
\end{equation}
Neutral vacua are realized for $u=0$. There are four types of neutral extrema (so also vacua) that have different symmetry properties:
\begin{itemize}
\item[(i)] The $EW\!s$ case with $u = v_D = v_S = 0$  corresponds to the EW symmetry. 
\item[ (ii)] Mixed vacuum $(M)$ with  $u = 0, v_S \not =0, v_D \not =0 $. There exist two charged Higgs particles $H^\pm$, a pseudoscalar Higgs $A$ and two CP-even Higgses $h$ and $H$, both of them could be SM-like.
\item[(iii)] Inert vacuum $(I_1)$ with  $u = v_D = 0, v_S\not =0$ is the only state that conserves the $D$-parity and assures the existence of a stable scalar particle - the DM candidate.
\item[(iv)] Inertlike vacuum $(I_2)$ with $u = v_S = 0, v_D\not =0$, which  spontaneously violates the $D$ symmetry by $v_D \not = 0$.
\end{itemize}
If $u \not = 0$ then the charged vacuum $(CB)$ is realized. This leads to the U(1)$_{QED}$ symmetry breaking and the appearance of a massive photon. Such a case is not realized in the nature today.

Mixed and charged minima cannot exist for the same values of the parameters of the potential $V$. On the other hand, minima of the inert-type ($I_1$ or $I_2$) can overlap one another and $M$ or $CB$ in the parameter space, e.g. in the  $(\lambda_4, \lambda_5)$ plane~\cite{Krawczyk:2009fb}.

\begin{figure}[h]
\centering
\subfloat[$R>1$]{\label{phasediaa}
\includegraphics[width=.3\textwidth]{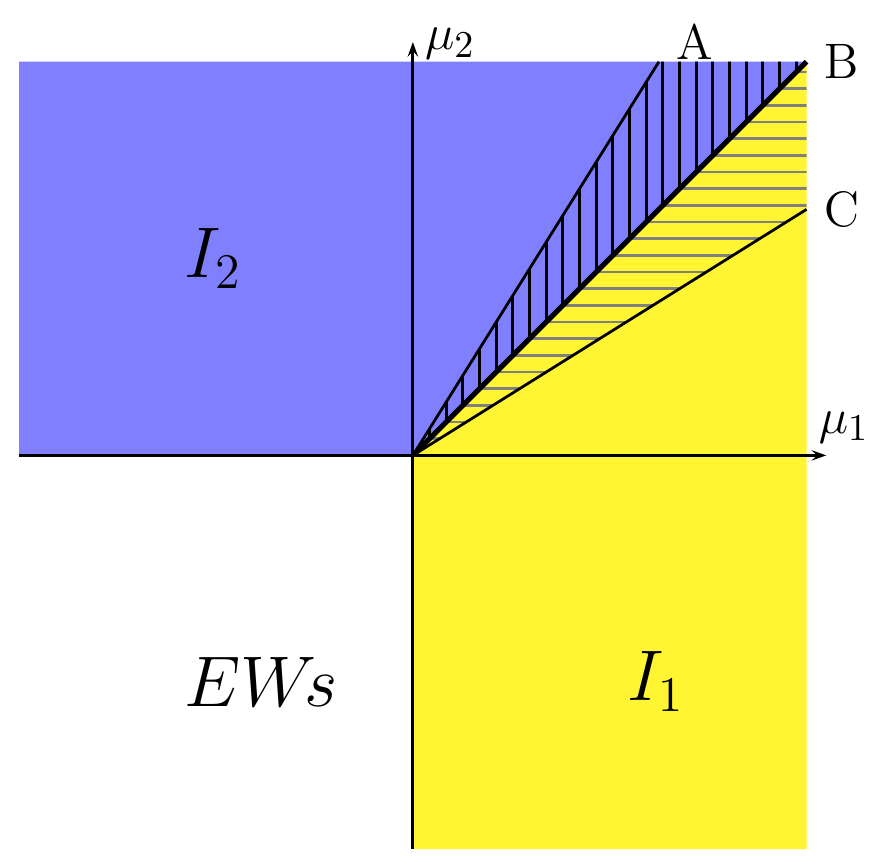}}
\subfloat[$1>R>0$]{
\includegraphics[width=.3\textwidth]{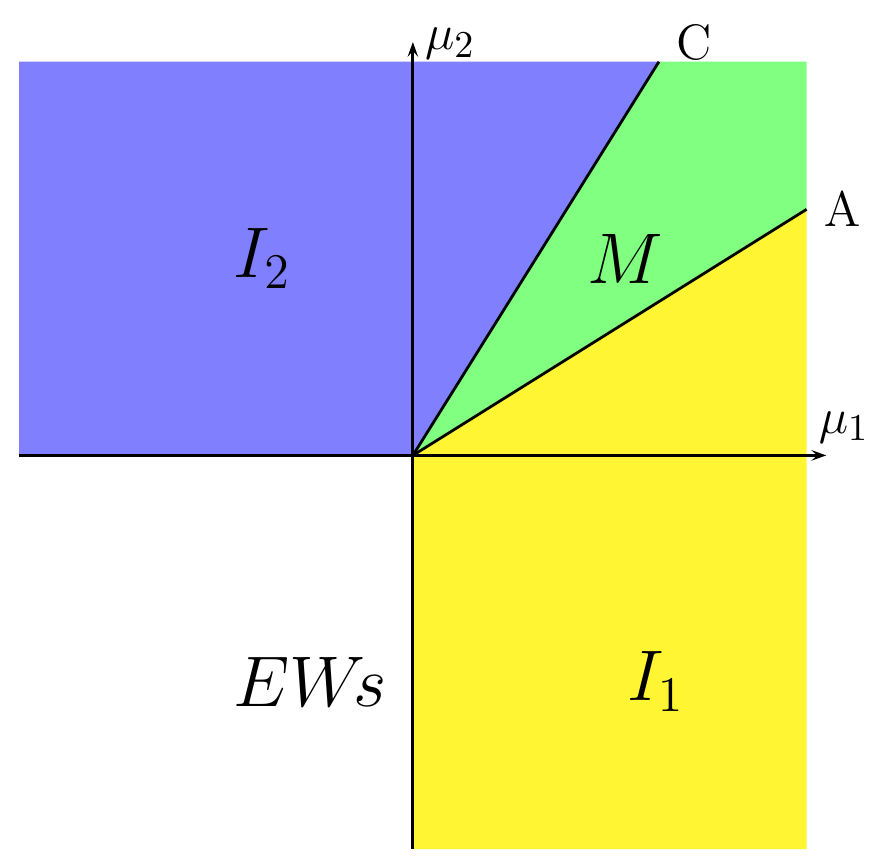}}
\subfloat[$0>R>-1$]{
\includegraphics[width=.3\textwidth]{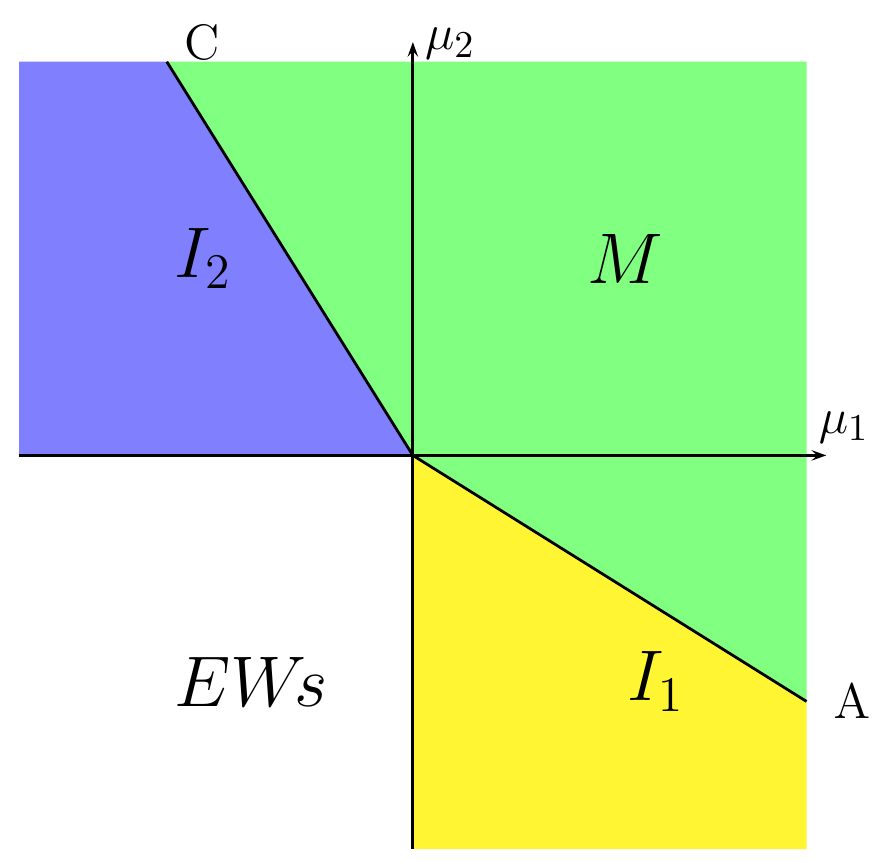}}

\caption{Phase diagrams  for the neutral vacua in the $D-$symmetric potential:
regions of $\mu_1, \mu_2$ where various neutral minima $EW\! s$ (white), $I_1$ (yellow/light gray), $I_2$ (blue/dark gray), $M$ (green/medium gray) can be realized are shown. The hatched region in the panel (a) corresponds to the coexistence of $I_1, I_2$ minima. } \label{phasedia}
\end{figure}

A stable vacuum exists only if 
\begin{equation}
\la>0,\quad\lb>0,\quad\lc+\sqrt{\la\lb}>0,\quad\lczp+\sqrt{\la\lb}>0\,\,\,\, (\lczp=\lambda_3+\lambda_4+\lambda_5), \label{stability}
\end{equation}
so that $$R=\fr{\lambda_{345}}{\sqrt{\lambda_1}\sqrt{\lambda_2}} > -1. $$ Extremum with the lowest energy corresponds to the global minimum, i.e. the vacuum. 
The vevs (\ref{dekomp_pol}) are related to the parameters of the potential through the extremum conditions and so the various values of $v_S, v_D, u$ can be  represented  on  the 
phase diagram $(\mu_1,\mu_2)$, where
\begin{displaymath}
\mu_1 =\frac{m_{11}^2}{\sqrt{\la}},\quad \mu_2=\frac{m_{22}^2}{\sqrt{\lb}}.
\end{displaymath}
Different regions of this parameter space correspond to the different vacua. 
There are three  regimes of parameter $R$ which correspond to very different phase patterns shown in Fig. \ref{phasedia}. There is a  possibility of unique coexistence of two inert-type minima,  as shown in Fig.~\ref{phasediaa}.

Analogous phase diagrams can be considered for $EW\!s, I_1, I_2, CB$ with $R \to R_3 = \lambda_3/\sqrt{\lambda_1\lambda_2}$ and $M \to CB$.

\section{IDM}\label{idm}
The Inert Doublet Model (IDM) is realized if   $I_1$ is the  vacuum state. Here only one doublet ($\phi_S$) is involved in the SSB and there is only one  SM-like Higgs boson $h$ - we assume that its mass is equal to 125 GeV.  The doublet $\phi_D$  has vev $= 0$  and Yukawa interaction is as in Model~I, since $\phi_D$ does not interact with fermions (it is inert). The $D$ symmetry is exact here and only the second doublet is $D$-odd. It contains four scalars $H,\,A,
H^\pm$, called also dark scalars ($D$-scalars). The lightest neutral scalar $H$ (or $A$) may play a role of the DM. We take  DM $=H$ (so $\lambda_{4}+\lambda_5<0$).

Masses of the scalar particles read:
\begin{displaymath}
M_{h}^2=\lambda_1v^2= m_{11}^2\,,\quad M_{H^\pm}^2=\fr{\lambda_3 v^2-m_{22}^2}{2}\,,\quad
M_{A}^2=M_{H^\pm}^2+\fr{\lambda_4-\lambda_5}{2}v^2\,,\quad M_{H}^2=
M_{H^\pm}^2+\fr{\lambda_4+\lambda_5}{2}v^2\,.
\end{displaymath}
with $v=246$ GeV. 

These masses and the physical couplings can be used to describe the properties of the~IDM. $\lambda_{345}$ is related to  triple and quartic couplings between the SM-like Higgs $h$ and the DM candidate $H$. $\lambda_2$ gives the quartic DM self-couplings, while $\lambda_3$ describes the Higgs particle interaction with charged scalars. These parameters are  subject to various  theoretical and experimental constraints (see e.g.~\cite{Cao:2007rm, Agrawal:2008xz, Gustafsson:2007pc, Dolle:2009fn, Dolle:2009ft, LopezHonorez:2006gr, Arina:2009um, Tytgat:2007cv, Honorez:2010re, Lundstrom:2008ai, Krawczyk:2009fb, Kanemura:1993, Akeroyd:2000, Swiezewska:2012, Gustafsson:2009, Gustafsson:2010}): \\

\paragraph{Perturbative unitarity constraints on self-couplings} Parameters $\lambda$ of $V$ are constrained by the condition: $|\Lambda_i|<8\pi$~\cite{Kanemura:1993,Akeroyd:2000,Swiezewska:2012}, where $\Lambda_i$ are the eigenvalues of the high-energy scattering matrix of the scalar sector. This leads to the following limits:
\begin{displaymath}
\lambda_{1,2}^{\textrm{max}} = 8.38, \quad \lambda_3 \in (-6.05, 16.53), \quad \lambda_{345} \in (-8.10,12.38).
\end{displaymath}
\paragraph{Constraints from existence of $I_1$ vacuum} Existence of the normal (Mixed) vacuum is equivalent to having positive scalars' masses squared (condition for the minimum). It is not enough in the case of the Inert vacuum, as  a coexistence of Inert and Inertlike minima is possible (see Fig.~\ref{phasediaa}). For the Inert state to be the global minimum  in addition the following  condition has to be fulfilled~\cite{Krawczyk:2010}:
\begin{equation}
\frac{m_{11}^2}{\sqrt{\la}}>\frac{m_{22}^2}{\sqrt{\lb}}.\label{inertvac}
\end{equation}
Knowing  the Higgs boson mass $M_h=125\g $ and using the unitarity limit $\lb^{\textrm{max}}=8.38$ one can derive the following  limit on $m_{22}^2$, that was not considered before~\cite{Swiezewska:2012}:
\begin{displaymath}\label{m22bound}
m_{22}^2\lesssim 9\cdot10^4\g^2.
\end{displaymath}

\paragraph{Electroweak precision tests} EWPT strongly constrain  physics beyond SM.  Values of $S$ and $T$ parameters are demanded to lie within $2\sigma$ ellipses in the $(S,T)$ plane, with the following central values~\cite{Nakamura:2010}: $S=0.03\pm0.09$, $T=0.07\pm0.08$, with correlation equal to 87\%. 

\paragraph{LEP II limits} 
The LEP II analysis excludes the region of masses where simultaneously: $M_{H} < 80$ GeV, $M_{A} < 100$ GeV and $M_A-M_H > 8$ GeV. For $M_A-M_H <8$ GeV the LEP I limit $M_{H} + M_{A} > M_Z$ applies \cite{Gustafsson:2009,Gustafsson:2010}.

\paragraph{$\lambda_2$ parameter}
Neither there exist, nor are expected in the near future, experimental limits  on quartic selfcoupling for dark scalars $\lambda_2$.   However, as we pointed out in \cite{Sokolowska:2011aa,Sokolowska:2011sb} there are some  indirect constraints for $\lambda_2$ that come from its relation to $\lambda_{345}$ through the vacuum stability constraints (\ref{stability}) and existence of $I_1$ vacuum (\ref{inertvac}).

\section{Relic density constraints}
\begin{figure}[h]
\centering
\subfloat[$M_{A} = M_{H} + 8$ GeV]{\label{rDMa}
\includegraphics[width=0.4\textwidth]{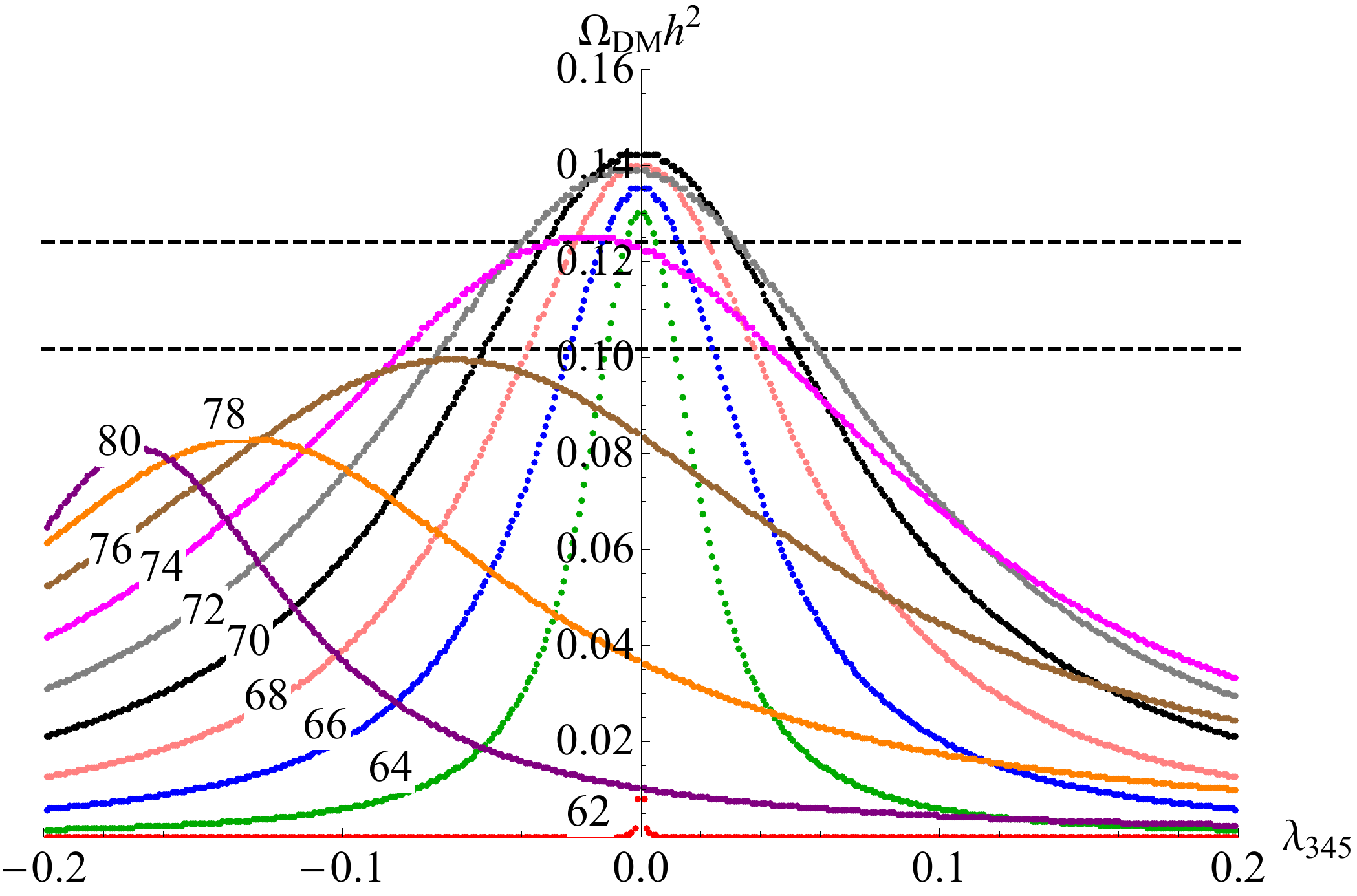}} \hspace{.5cm}
\subfloat[$M_{A} = M_{H} + 50$ GeV]{\label{rDMb}
\includegraphics[width=0.4\textwidth]{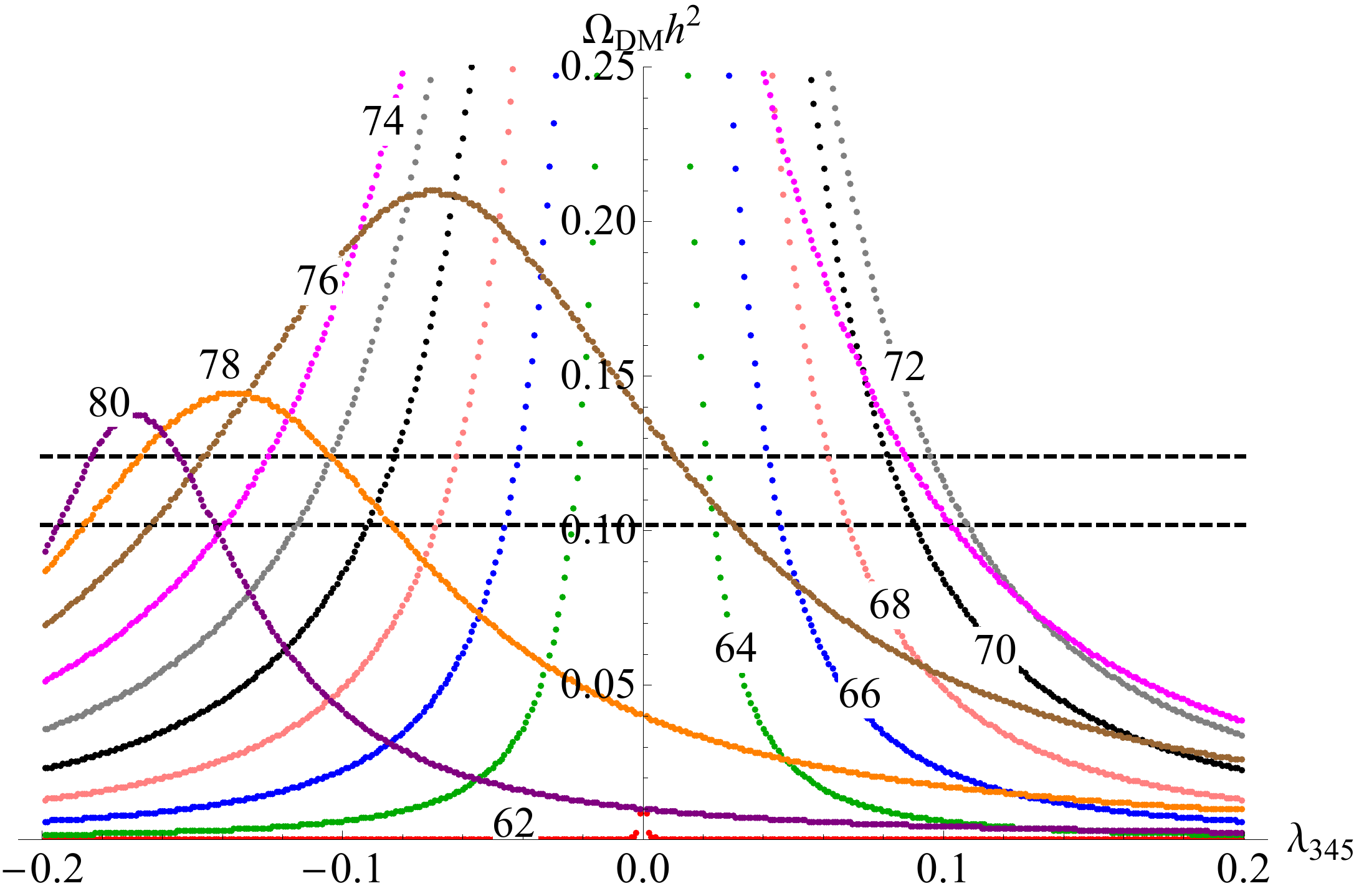}}
\caption{$(\lambda_{345},\Omega_{DM} h^2)$ for various values of $M_H$ with (a) and without (b) coannihilation. Horizontal lines denote 3$\sigma$ WMAP limit: $0.1018<\Omega_{DM} h^2<0.1234$. For $\Omega_{DM} h^2<0.1018$ $H$ from IDM does not constitute the whole DM in the Universe. Region of $\Omega_{DM} h^2>0.1234$ is excluded.
Numbers on the  curves denote the DM mass $M_H$ (GeV).
We set $M_h = 125$ GeV and $M_{H^\pm} = M_H + 50$ GeV.}\label{rDM}
\end{figure}

IDM provides a good DM candidate ($H$) in agreement with the data on relic density $\Omega_{DM} h^2$, $\Omega_{DM}h^2=0.1126 \pm 0.0036$ \cite{Beringer:2012},  in three regions of $M_H$ \cite{Cao:2007rm, Barbieri:2006dq, Gustafsson:2007pc, Dolle:2009fn, Dolle:2009ft, LopezHonorez:2006gr, Arina:2009um, Tytgat:2007cv, Honorez:2010re,LopezHonorez:2010tb,Sokolowska:2011aa,Sokolowska:2011sb}:
\begin{itemize}
\item light DM particles with mass below $10 \textrm{ GeV}$. Other dark particles are much heavier than DM and so this scenario mimics the behavior of the singlet DM model.
\item  medium mass regime of $50-150 \textrm{ GeV}$ with two distinctive regions: with and without coannihilation of $H$ with the  neutral $D$-odd particle~$A$.
\item heavy DM of mass larger than $500 \textrm{ GeV}$. In this region all dark particles have almost degenerate masses and coannihilation processes between all dark particles are crucial for the agreement with the measured  DM relic density.
\end{itemize}
 The relic density data can be used to constrain   $\lambda_{345}$ coupling for  chosen values of masses of $H$ and other scalars \cite{Dolle:2009fn,LopezHonorez:2006gr}.
In Fig.~\ref{rDM} the $\Omega_{DM} h^2$ as a function of $\lambda_{345}$ for various values of the DM mass is shown. Horizontal lines correspond to the 3$\sigma$ limit $0.1018<\Omega_{DM}h^2<0.1234$. 

For both cases, i.e. (a)  with  and (b) without  coannihilation of $H$ and $A$, shown respectively in Fig. \ref{rDMa}  and  Fig. \ref{rDMb},   the allowed region of $\lambda_{345}$ is symmetric around zero for masses $M_H \lesssim 72$~GeV. Usually small values of $\lambda_{345}$ are excluded due to a non efficient DM annihilation. In the case (a) allowed values of $\lambda_{345}$ are smaller than in the case (b), as the process $HH \to \bar{f}{f}$ with the cross-section $\sigma \sim \lambda_{345}^2$ does not have to be that efficient to provide the proper relic density value. As  mass increases, the region of proper relic density shifts towards the negative values of $\lambda_{345}$, which is due to opening of the annihilation channels into the gauge bosons final state and interference of processes $HH \to h \to VV$ and $HH \to VV$. 

As it will be shown in the next section,  in  the IDM the enhancement of the $\gamma \gamma$ decay rate for the 125 GeV Higgs  is possible for $\lambda_3<0$. If $H$ is the lightest dark particle (DM candidate) then also $\lambda_{4}+\lambda_5 <0$ and so we get $\lambda_{345} <0$ for such case.
For a recent analysis of the DM properties in the IDM and its relation to the two-photon decay rate of the Higgs boson see also Ref.~\cite{Stal:2013}.

It is worth noticing that the relic density as all other experimental data does not limit   the $\lambda_2$ parameter.

\section{The two-photon decay rate of the Higgs boson}\label{LHC}

\begin{figure}[h]
\centering
\subfloat[Values of $\rg$ allowed by the theoretical and experimental constraints  as a function of the DM mass $M_H$ for $-2\cdot10^6\g^2\leqslant m_{22}^2\leqslant 9\cdot10^4\g^2$.\label{mDM}]{
\includegraphics[width=0.4\textwidth]{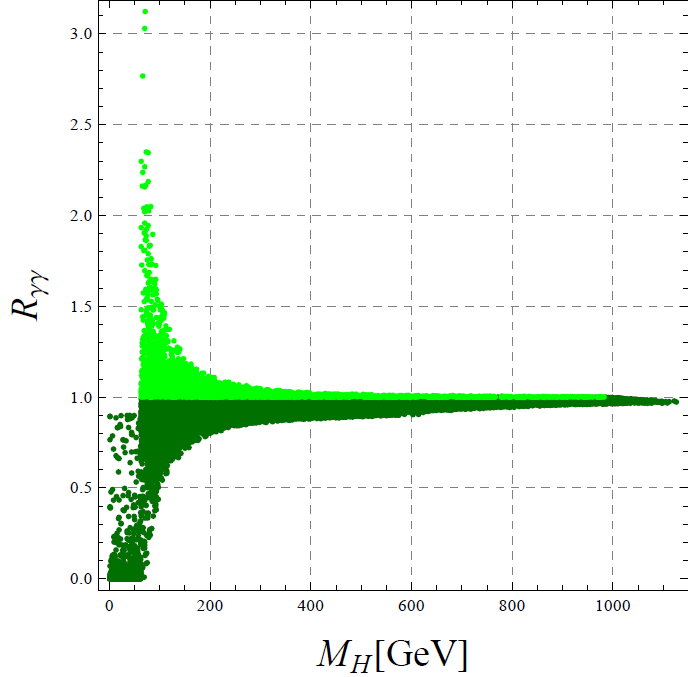}} \hspace{.5cm}
\subfloat[Region allowed by the theoretical and experimental constraints in the $(m_{22}^2,\m)$ plane. The curves correspond to the fixed values of $\rg$ (for the invisible channels closed).\label{mHp}]{
\includegraphics[width=0.4\textwidth]{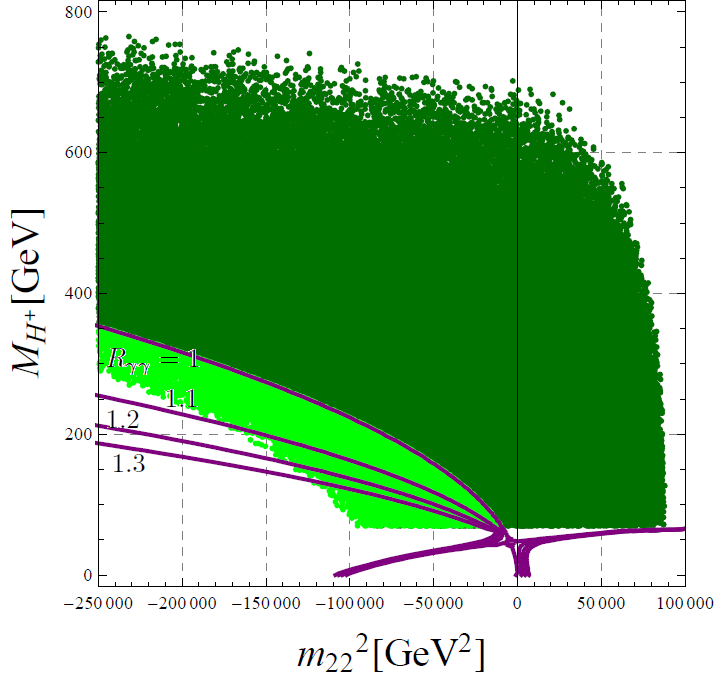}}
\caption{Results on $\rg$ for IDM, points with $\rg<1$ ($\rg>1$) are displayed in dark green/gray (light green/gray).\label{m}}
\end{figure}

The 
 two-photon decay rate of the Higgs boson in the IDM reads:
\begin{displaymath}\label{rgg}
R_{\gamma \gamma}:=\frac{\sigma(pp\to h\to \gamma\gamma)^{\textrm{IDM}}}{\sigma(pp\to h\to \gamma\gamma)^{\textrm  {SM}}}
\approx\frac{\textrm{Br}(h\to\gamma\gamma)^{\textrm {IDM}}}{\textrm{Br}(h\to\gamma\gamma)^{\textrm {SM}}},
\end{displaymath}
where the fact that the main production channel is the gluon fusion and that the Higgs particle in IDM is SM-like (in the sense defined before) was taken into account.
In the IDM there are two possible sources of deviation from $\rg=1$:
\begin{itemize} 
\item A contribution from the charged scalar loop to the partial decay width $\Gamma(h\to \gamma\gamma)^{\textrm{IDM}}$  ~\cite{Djouadi:2005, Djouadi:2005sm, Ma:2007, Posch:2010, Arhrib:2012}:
$$
\Gamma(h\to\gamma\gamma)^{\textrm{IDM}}=\frac{G_F\alpha^2M_h^3}{128\sqrt{2}\pi^3}\left | \mathcal{M}^{\textrm{SM}}+\delta\mathcal{M}^{\textrm{IDM}}\right |^2,
$$
where $\mathcal{M}^{\textrm{SM}}$ is the amplitude present in the SM and $\delta\mathcal{M}^{\textrm{IDM}}=\frac{2\m^2+m_{22}^2}{2\m^2}A_0\bigg(\frac{4\m^2}{M_h^2}\bigg)$, with $2\m^2+m_{22}^2=\lc v^2$ (the definition of the function  $A_0$ can be found in Ref.~\cite{Djouadi:2005, Djouadi:2005sm}).
The charged scalar contribution can interfere either constructively or destructively with the SM contribution, this way increasing or decreasing the decay rate.
\item  The invisible decays: $h\to HH$ and $h\to AA$ can augment the total decay width $\Gamma^{\textrm{IDM}}(h)$  with respect to the SM case.  
\end{itemize}
Performing a  scan of the parameter space~\cite{rg:2012}, taking into account  the theoretical and experimental constraints presented  in Sec.~\ref{idm}, we found the regions where $\rg $>1, with the maximal value of $\rg$  around $3.4$.
Fig.~\ref{mDM} shows values of $\rg$  as a function of $M_H$ -  it is clear that  $\rg>1$ is not possible for $M_H<M_h/2\sim62.5$ GeV. It means that if the invisible channels are open, they suppress the charged scalar loop effect and enhancing $\rg$ is impossible.

Fig.~\ref{mHp} shows the viable parameter space in the $(m_{22}^2,\m)$ plane together with the curves corresponding to constant values of $\rg$ (calculated for the case with invisible decay channels closed). It can be seen that than the $\rg$ enhancement is possible only for:
$$
m_{22}^2<-9.8\cdot10^3\g^2,
$$
which is equivalent to the bound $\lc<0$,  and consequently $\lambda_{345}<0$, see also Ref.~\cite{Arhrib:2012}. We found that $\rg>1$ can be achieved for any value of $\m$. However, if larger value of $\rg$ is demanded, then allowed values of $\m$ are constrained. For example, for $\rg>1.2$ we get the following bounds on $\m$ and $M_H$ (as $M_H<\m$): 
$$
\ba{rcccl}
62.5\g&<&M_H&<&154 \g,\\[-2pt]
70\g&<&\m&<&154 \g.\\
\ea
$$

\section{Evolution of the Universe towards the Inert phase}
We assume that today we live in the  $I_1$ vacuum, however in the past the vacuum  could have been different. We first consider  the simplest description of the evolution of the Universe with the thermal $T^2$ corrections. In this approximation, valid for high temperatures $T\gg M$,  $\lambda$ parameters are fixed and only mass parameters vary with 
temperature as follows:
\begin{equation}
 m_{ii}^2(T)=m_{ii}^2-c_iT^2\,\,(i=1,2),\nonumber
\end{equation}
where
\begin{displaymath}
c_1=\fr{3\lambda_1+2\lambda_3+\lambda_4}{6}+\fr{3g^2+g^{\prime 2}}{8}+\fr{g_t^2+g_b^2}{2}, \quad
c_2=\fr{3\lambda_2+2\lambda_3+\lambda_4}{6}+\fr{3g^2+g^{\prime 2}}{8}.		
\end{displaymath}
Here $g, g\rq{}$ are the EW gauge couplings, while  $g_t,g_b$ are the  SM Yukawa couplings \cite{Krawczyk:2010,gil1}.
From the positivity conditions and the positivity of the vevs squared it follows that $c_1 + c_2>0$.

The quadratic coefficients of $V$ change with temperature $T$ and so the ground state of the  potential may change. There exist three types of evolution of the Universe from the EW symmetric state to the $I_1$ phase in one, two or three steps:  $EW\!s \to I_1$,  $EW\!s \to I_2 \to I_1$  or  $EW\!s \to I_2 \to M \to I_1$. In general, in the considered $T^2$ approximation, phase transitions are of the second order with the exception of the $I_2 \to I_1$ transition in the two-stage evolution. This is a transition between two degenerate minima $I_2$ and $I_1$ and it can be realized only if $R>1$. 
Notice, that the DM exists only in the $D$-symmetric state $I_1$ and so it may appear later during the evolution, if the Universe goes through the two- or three-stage  sequences.

\begin{figure}[h]
\centering
\subfloat[$R<0$: the only sequence  that leads to restoration of EW symmetry in the past.\label{rR31}]{
\includegraphics[width=0.3\textwidth]{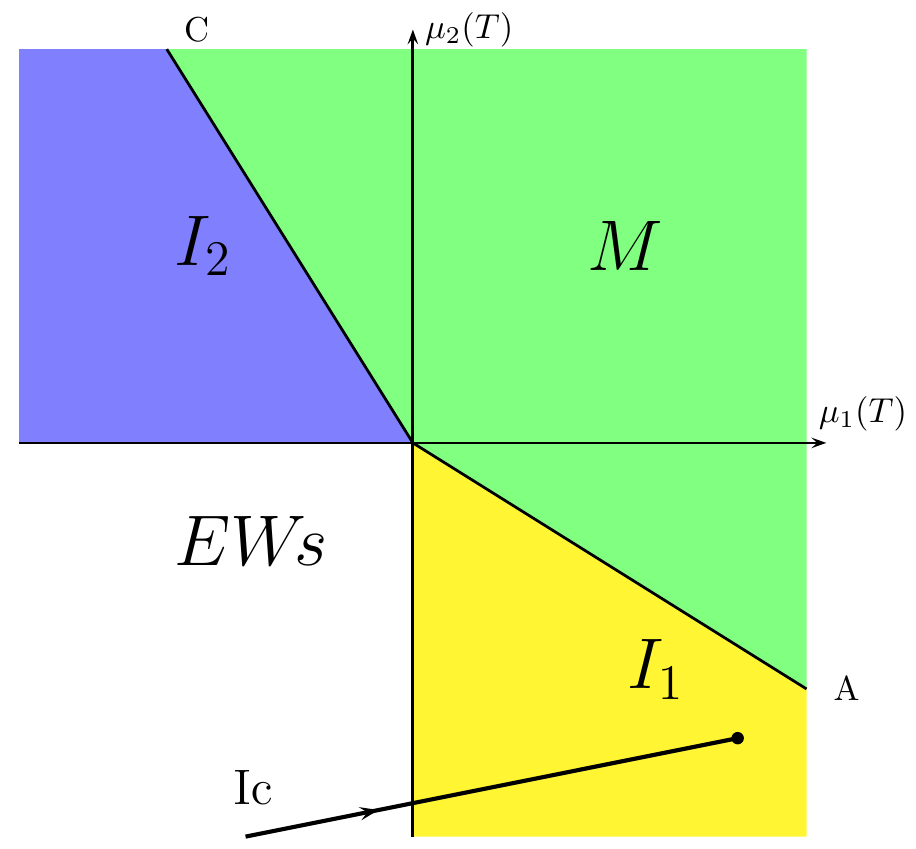}} \hspace{.2cm}
\subfloat[$R<0$: possible sequences with non-restoration of EW symmetry.\label{rR32}]{
\includegraphics[width=0.3\textwidth]{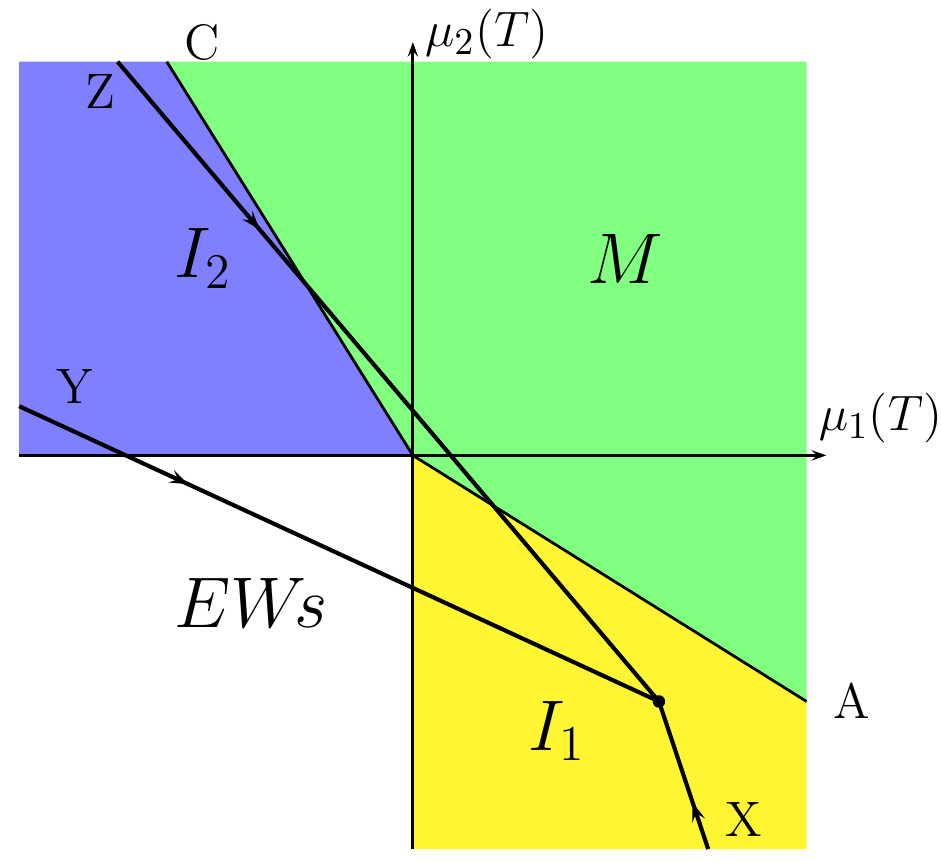}} \hspace{.2cm}
\subfloat[$R_3<0$: transition through the charged vacuum.\label{rR33}]{
\includegraphics[width=0.3\textwidth]{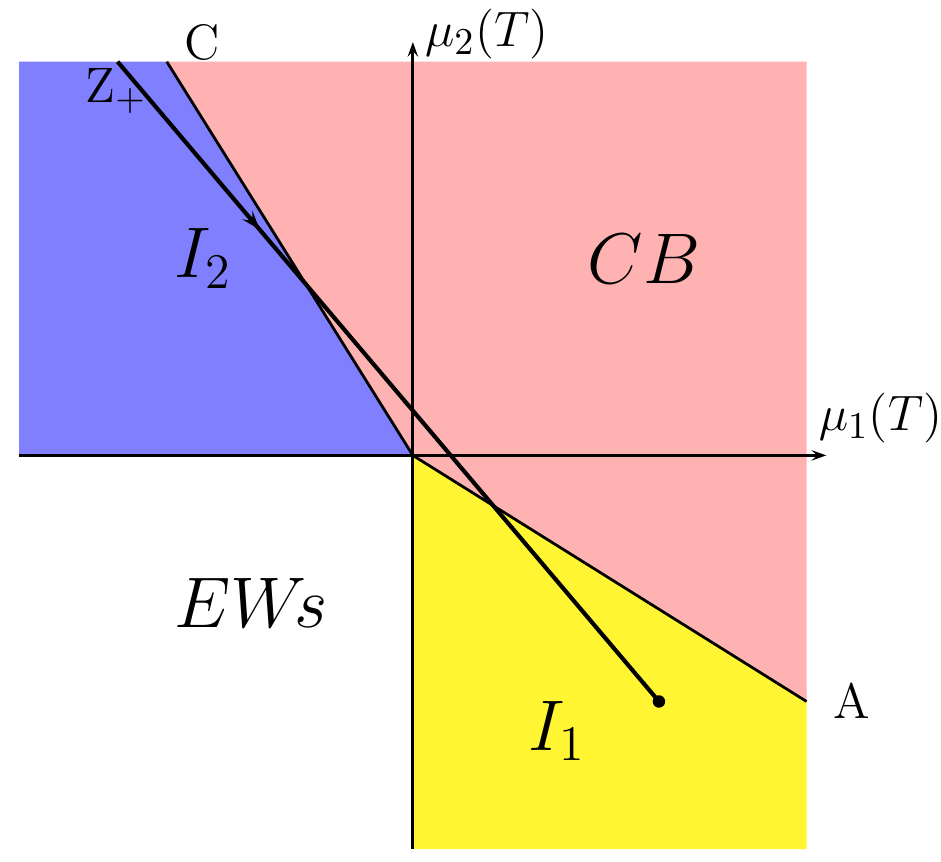}}
\caption{Possible sequences of phase transitions for $R<0$.}
\label{rR3}
\end{figure}

If $R<0$ only one sequence that corresponds to the restoration of EW symmetry in the past is possible, namely  $EW\!s \to I_1$ (Fig.~\ref{rR31}). It can be realized when $R_{\gamma \gamma} > 1$, which is suggested  by the recent LHC data, as discussed above. In the other scenarios for $R<0$ in the initial state of the Universe there is no EW symmetry (Fig. \ref{rR32}). There is a~possibility of a temporary restoration of the EW symmetric state (scenario Y), in other cases the EW symmetric state never existed.

For a certain parameter range  there is a theoretical possibility of having a charged vacuum in the past (Fig.~\ref{rR33})~\cite{Krawczyk:2010}. This type of evolution is equivalent to having IDM with the charged DM particle \cite{Krawczyk:2010}. 
Current model-independent bounds require the charged DM  to be heavier than 100 TeV \cite{Chuzhoy:2008zy}. Heavy charged scalar may appear in the model  without breaking the perturbative unitarity conditions for very negative  $m_{22}^2$, i.e.  large |$m_{22}^2$|. However, the sequence $Z_+$ requires that $c_2<0$ and $|c_2|/c_1>|m_{22}^2|/m_{11}^2\gtrsim (10^5\div 10^6)$. This contradicts the requirement $c_1>-c_2$ based on the positivity condition and invalidates the possibility of passing through the charge breaking phase during the evolution \cite{Krawczyk:2010}. 

\section{Temperature evolution beyond $T^2$ approximation}
The $T^2$ approximation, used in the previous section, simplifies the discussion so that all phase sequences, that are in principle possible, can be recognized and analyzed in details.   However, this approach has obvious limitations, since some sequences are realized at relatively low temperatures, where the  $T^2$ approximation is not valid.
The analysis  of the evolution of the Universe to the Inert phase  going beyond this approximation was performed in \cite{gil1}, where the effective potential approach was used under the assumption  that only neutral vacua (phases) were realized  in the past. It was found, as expected, that some of the second-order phase transitions change their character to the first-order.  The analysis which  focused on the strength of the first-order phase transition  needed for baryogenesis was performed,  based on the one-loop effective potential at $T=0$ (Coleman-Weinberg term) and temperature dependent effective potential at $T \not = 0$ (including sum of the ring diagrams) \cite{Gil:2012ya}.  Of course, IDM as a model with the CP-conserving  potential can not serve alone as a realistic model for baryogenesis and  an additional  source of CP violation has to be included.  Nevertheless, one can examine
whether the model parameters, for which the right density of relic particles is predicted, are compatible with the strong first-order phase transition that could allow the electroweak baryogenesis. 
 
The strong first-order transition is realized  if for  the critical temperature $T_{EW}$ we have  $v(T_{EW})/T_{EW}>1$. We found that this condition is fulfilled in the $M_h$=125 GeV case with medium DM mass $M_H \sim 60$ GeV and relatively  heavy dark scalars $A$ and $H^\pm$, with masses $\sim 300-400$ GeV  (we took them to be degenerate for simplicity). The important parameter is $\lambda_{345}$ - the coupling of the Higgs particle $h$ to the DM ($hHH$), while the parameter $\lambda_2$ does not play any role in the analysis (we fix it to be equal to 0.2).
For the mass of the DM equal to 65~GeV  we obtained allowed mass region of heavier dark scalars  between 275 and 380 GeV ($M_A = M_{H^\pm}$), and in order to be in agreement with XENON-100 data  $|\lambda_{345}|$ should be smaller than 0.1. This is in agreement with similar analyses~\cite{Borah:2012pu,Chowdhury:2011ga}, see also papers for 2HDM (not IDM) \cite{Cline:2011mm,Kozhushko:2011ea,Kanemura:2004mg,Kanemura:2004ch}.  It is worth noticing that these regions of parameters correspond to the one-stage evolution of the Universe, and the negative $\lambda_{345}$ corresponds to the enhanced $h \to \gamma \gamma$ decay rate. 

\section{Summary and outlook}
2HDM is a great laboratory of BSM. A distinguished version of 2HDM is the  Inert Doublet Model  with the  SM-like Higgs sector and the  dark sector with a good DM candidate. Recent  LHC data with the SM-like Higgs signal, with a possible enhancement in the $\gamma\gamma$ channel,  put strong constraints on masses of DM and other dark scalars, as well as the self-couplings, especially $\lambda_{345}$. A striking conclusion is that an  enhanced $h \to \gamma \gamma$  rate excludes a light DM particle  and  sheds light on  the evolution of the Universe, suggesting only  one phase transition from the  $EW\!s$  to  the today's  Inert phase  with the right DM relict density. 

The precise measurements of the SM-like Higgs properties are crucial not only to establish details of the EWSB mechanism, but also to understand  the composition of  the Dark Matter. The future linear collider in the $e^+e^-$ and $\gamma \gamma, e \gamma$ options may indeed lead to further progress in these directions \cite{DeRoeck:2009id,Djouadi:2007ik}. It is known that for example the partial width $\Gamma(H_{SM}\to\gamma\gamma)$  can be measured at  linear collider with $\sim 2 \%$   precision, by combining the 2\% measurement of the Higgs particle production in $\gamma \gamma$ collision decaying into $b\bar b$,  $\Gamma(H_{SM}\to\gamma\gamma)Br(H_{SM}\to b \bar b)$, with the measurement of the $Br(H_{SM}\to b \bar b)$ to be performed with 1\% precision at  the $e^+e^-$ option of ILC at 500 GeV \cite{Niezurawski:2005tw}.   Some  analysis on IDM at the ILC  has just appeared \cite{Aoki:2013lhm}.     
\begin{acknowledgments}
MK is grateful to the Organizers for this very interesting Workshop and exceptional hospitality. She is thankful for the discussion with  A.~Arhrib and I.~Ginzburg on $hZ\gamma$ coupling. We thank   P.~Chankowski and  G.~Gil for a~fruitful collaboration.
The work was supported in part by a grant  NCN OPUS 2012/05/B/ST2/03306 (2012-2016). 
\end{acknowledgments}

\bibliography{biblio1}

\end{document}